\documentclass[10pt, conference, letterpaper]{IEEEtran}
\usepackage[T1]{fontenc}
\usepackage[utf8]{inputenc}
\usepackage{microtype}
\usepackage{pifont}

\makeatletter
\let\@ORGmakecaption\@makecaption
\long\def\@makecaption#1#2{\@ORGmakecaption{#1}{#2}\vskip\belowcaptionskip\relax}
\makeatother

\usepackage{amsmath,amssymb}

\usepackage{blkarray}
\usepackage{bm}
\usepackage{booktabs,multirow, array, enumitem}
\usepackage{tabularx}
\usepackage{graphicx}
\usepackage[caption=false,font=footnotesize,subrefformat=parens,labelformat=parens]{subfig}
\usepackage{capt-of}
\usepackage{xcolor}
\usepackage{siunitx}
\usepackage[most,breakable]{tcolorbox}
\usepackage{xurl}
\usepackage[noadjust]{cite}
\usepackage{xspace}
\usepackage[printonlyused]{acronym}
\usepackage{listings}
\usepackage[ruled,vlined,linesnumbered]{algorithm2e}
\SetAlgoSkip{} 
\usepackage{float}
\usepackage{balance}
\usepackage{tikz}

\newcommand\publishedtext{%
  \footnotesize This is the author's accepted version of the article. The final version published by IEEE is:
  M. Jabbari, A. Duttagupta, C. Fiandrino, L. Bonati, S. D'Oro, M. Polese, M. Fiore, and T. Melodia,
  ``\toolname: Symbolic Interpretability for Anticipatory Deep Reinforcement Learning in Network Control,''
  IEEE INFOCOM 2026, doi: TBD.%
}
\newcommand\copyrighttext{%
  \footnotesize \textcopyright 2026 IEEE. Personal use of this material is permitted.
  Permission from IEEE must be obtained for all other uses, in any current or future
  media, including reprinting/republishing this material for advertising or promotional
  purposes, creating new collective works, for resale or redistribution to servers or
  lists, or reuse of any copyrighted component of this work in other works.%
}

\newcommand\copyrightnotice{%
\begin{tikzpicture}[remember picture,overlay]
\node[anchor=north,yshift=0pt] at (current page.north)
  {\fbox{\parbox{\dimexpr\textwidth-\fboxsep-\fboxrule\relax}{\publishedtext}}};
\node[anchor=south,yshift=10pt] at (current page.south)
  {\fbox{\parbox{\dimexpr\textwidth-\fboxsep-\fboxrule\relax}{\copyrighttext}}};
\end{tikzpicture}%
}

\newcommand{\toolname}{\textsc{SIA}\xspace}
\newcommand{\toolnamex}{\textsc{SIA}}

\newcommand{\AoneR}{A1-R\xspace}
\newcommand{\AoneP}{A1-P\xspace}
\newcommand{\AonePSIA}{A1-P+SIA\xspace}

\newcommand{\AtwoR}{A2-R\xspace}
\newcommand{\AtwoP}{A2-P\xspace}
\newcommand{\AtwoRSIA}{A2-R+SIA\xspace}

\newcommand{\AthreeR}{A3-R\xspace}
\newcommand{\AthreeRSIA}{A3-R+SIA\xspace}

\DeclareTextFontCommand{\mytexttt}{\ttfamily\hyphenchar\font=45\relax}
\newcommand\txbrate{\relax\ifmmode\mathtt{tx\_bit\-rate}\else\mytexttt{tx\_bit\-rate}\fi\xspace}
\newcommand\txpkts{\relax\ifmmode\mathtt{tx\_packets}\else\mytexttt{tx\_packets}\fi\xspace}
\newcommand\dlbuff{\relax\ifmmode\mathtt{DWL\_buffer\_size}\else\mytexttt{DWL\_buffer\_size}\fi\xspace}

\SetCommentSty{mycommfont}

\usepackage[printonlyused]{acronym}
\acrodef{ai}[AI]{Artificial Intelligence}
\acrodef{ml}[ML]{Machine Learning}
\acrodef{dl}[DL]{Deep Learning}
\acrodef{aml}[AML]{Adversarial Machine Learning}
\acrodef{xai}[XAI]{EXplainable Artificial Intelligence}
\acrodef{nn}[NN]{Neural Networks}
\acrodef{dnn}[DNN]{Deep Neural Networks}
\acrodef{rnn}[RNN]{Recurrent Neural Networks}
\acrodef{lstm}[LSTM]{Long-Short Term Memory}
\acrodef{gnn}[GNN]{Graph Neural Networks}
\acrodef{shap}[SHAP]{SHapely Additive exPlanations}
\acrodef{lime}[LIME]{Local Interpretable Model-agnostic Explanations}
\acrodef{xgb}[XGBoost]{Extreme Gradient Boosting}
\acrodef{lrp}[LRP]{LayeR-wise backPropagation}
\acrodef{nlp}[NLP]{Natural Language Processing}
\acrodef{tcn}[TCN]{temporal convolutional network}
\acrodef{gpr}[GPR]{Gaussian process regression}
\acrodef{dt}[DT]{Decision Tree}
\acrodefplural{DT}[DTs]{Decision Trees}
\acrodef{sai}[Symbolic AI]{Symbolic Artificial Intelligence}
\acrodef{mlp}[MLP]{Multi-layer Perceptron}

\acrodef{xrl}[XRL]{EXplainable Reinforcement Learning}
\acrodef{rl}[RL]{Reinforcement Learning}
\acrodef{drl}[DRL]{Deep Reinforcement Learning}
\acrodef{crl}[CRL]{Casual Reinforcement Learning}
\acrodef{dqn}[DQN]{Deep Q-Network}
\acrodef{ppo}[PPO]{Proximal Policy Optimization}
\acrodef{a3c}[A3C]{Asynchronous Advantage Actor-Critic}
\acrodef{fol}[FOL]{First-Order Logic}
\acrodef{mdp}[MDP]{Markov Decision Process}
\acrodef{sac}[SAC]{Soft Actor-Critic}
\acrodef{psr}[PSR]{Predictive State Representation}
\acrodef{mpc}[MPC]{Model Predictive Control}

\acrodef{fgsm}[FGSM]{Fast Gradient Sign Method}
\acrodef{bim}[BIM]{Basic Iterative Method}
\acrodef{mae}[MAE]{Mean Absolute Error}
\acrodef{rmse}[RMSE]{Root Mean Square Error}
\acrodef{mape}[MAPE]{Mean Absolute Percentage Error}

\acrodef{5g}[5G]{5th Generation}
\acrodef{6g}[6G]{6th Generation}
\acrodef{qos}[QoS]{Quality of Service}
\acrodef{sla}[SLA]{Service Level Agreement}
\acrodef{isp}[ISP]{Internet Service Providers}
\acrodef{bs}[BS]{Base Station}
\acrodefplural{BS}[BSs]{Base Stations}
\acrodef{qoe}[QoE]{Quality of Experience}
\acrodef{gnb}[gNB]{next Generation Node B}
\acrodefplural{gNB}[gNBs]{next Generation Node Bs}
\acrodef{enb}[eNB]{evolved Node B}
\acrodefplural{eNB}[eNBs]{evolved Node Bs}
\acrodef{ue}[UE]{User Equipment}
\acrodef{prb}[PRB]{Physical Resource Block}
\acrodefplural{PRB}[PRBs]{Physical Resource Blocks}
\acrodef{mcs}[MCS]{Modulation and Coding Scheme}
\acrodef{tti}[TTI]{Transmission Time Interval}
\acrodef{rnti}[RNTI]{Radio Network Temporary Identifier}
\acrodefplural{RNTI}[RNTIs]{Radio Network Temporary Identifiers}
\acrodef{tbs}[TBS]{Tranport Block Size}
\acrodef{rrc}[RRC]{Radio Resource Control}
\acrodef{ran}[RAN]{Radio Access Network}
\acrodef{lmf}[LMF]{Location Management Function}
\acrodef{amf}[AMF]{Access and Mobility Function}
\acrodef{vnf}[VNF]{Virtual Network Function}
\acrodef{embb}[eMBB]{enhanced Mobile BroadBand}
\acrodef{mmtc}[mMTC]{massive Machine-type Communication}
\acrodef{urllc}[URLLC]{Ultra-Reliable Low-Latency Communication}
\acrodef{v2x}[V2X]{Vehicle-to-Everything}
\acrodef{jmls}[JMLS]{Jump Markov Linear Systems}
\acrodef{mimo}[MIMO]{Multiple-Input Multiple-Output}
\acrodef{oran}[O-RAN]{Open Radio Access Network}
\acrodef{csi}[CSI]{Channel State Information}

\acrodef{ias}[IAS]{Intent-based Action Steering}
\acrodef{as}[AS]{Action-Steering}
\acrodef{kpi}[KPI]{Key Performance Indicator}
\acrodefplural{KPI}[KPIs]{Key Performance Indicators}
\acrodefplural{KPI}[KPIs]{Key Performance Indicators}
\acrodef{lfs}[LFS]{Learning Future Representation with Synthetic observations}
\acrodef{farl}[FARL]{Forecast-Augmented Reinforcement Learning}
\acrodef{ddtu}[DDTU]{Delta Data Transmitted for User}
\acrodef{kg}[KG]{Knowledge Graph}
\acrodef{is}[IS]{Influence Score}
\acrodef{mi}[MI]{Mutual Information}
\acrodef{kl}[KL]{Kullback-Leibler}

\acrodef{llm}[LLM]{Large Language Model}
\acrodefplural{LLM}[LLMs]{Large Language Models}
\acrodef{chkp}[CHKP]{Checkpoint}
\acrodefplural{CHKP}[CHKPs]{Checkpoints}
\acrodef{darpa}[DARPA]{Defense Advanced Research Projects Agency}
\acrodef{sota}[SOTA]{State-of-the-Art}
\acrodef{llm}[LLM]{Large Language Model}

\acrodef{dtu}[DTU]{Data Transmitted of User}
\acrodef{mase}[MASE]{Maximum Available Spectral Efficiency}
\acrodef{ug}[G]{User Group Label}
\acrodef{los}[LoS]{Line of Sight}
\acrodef{nlos}[NLoS]{Non-Line of Sight}
\acrodef{jfi}[JFI]{Jain-Fairness-Index}

\acrodef{abr}[ABR]{Adaptive Bitrate Streaming}

\newfloat{lstfloat}{htbp}{lop}
\floatname{lstfloat}{Listing}

\definecolor{siared}{rgb}{.647,.129,.149}
\definecolor{siagreen}{rgb}{0,0.53,0}
\convertcolorspec{cmyk}{0.94,0.53,0,0}{rgb}\siablue
\definecolor{siablue}{rgb}{\siablue}

\begin{document}

\bstctlcite{IEEEexample:BSTcontrol}

\title{
    \toolname: Symbolic Interpretability for Anticipatory Deep Reinforcement Learning in Network Control
    \vspace{-1ex}
}

\author{
    \IEEEauthorblockN{
        MohammadErfan Jabbari$^{*\dagger}$, 
        Abhishek Duttagupta$^{*\dagger}$, 
        Claudio Fiandrino$^{*}$, 
        Leonardo Bonati$^{\S}$,\\
        Salvatore D'Oro$^{\S}$, 
        Michele Polese$^{\S}$, 
        Marco Fiore$^{*}$, 
        and Tommaso Melodia$^{\S}$\\
        $^{*}$IMDEA Networks Institute, Spain, Email: \{name.surname\}@networks.imdea.org\\
        $^{\S}$Northeastern University, Boston, USA, Email: \{l.bonati, s.doro, m.polese, t.melodia\}@northeastern.edu\\
        $^{\dagger}$Universidad Carlos III de Madrid, Spain\vspace*{-10pt}
    }
\vspace{-5ex}%
}

\maketitle

% Added for the copyright
\copyrightnotice

% \begingroup
% \renewcommand\thefootnote{$\Diamond$}
% \footnotetext{These authors contributed equally to this work.}
% \endgroup

\begin{abstract}
\ac{drl} promises adaptive control for future mobile networks but conventional agents remain reactive: they act on past and current measurements and cannot leverage short-term forecasts of exogenous \acp{kpi} such as bandwidth. Augmenting agents with predictions can overcome this temporal myopia, yet uptake in networking is scarce because forecast-aware agents act as closed-boxes; operators cannot tell whether predictions guide decisions or justify the added complexity. We propose SIA, the first interpreter that exposes in real time how forecast- augmented DRL agents operate. SIA fuses Symbolic AI abstractions with per-\ac{kpi} Knowledge Graphs to produce explanations, and includes a new \ac{is} metric. SIA achieves sub-millisecond speed, over $200\times$ faster than existing \ac{xai} methods. We evaluate SIA on three diverse networking use cases, uncovering hidden issues, including temporal misalignment in forecast integration and reward-design biases that trigger counter-productive policies. These insights enable targeted fixes: a redesigned agent achieves a 9\% higher average bitrate in video streaming, and SIA's online Action-Refinement module improves RAN-slicing reward by 25\% without retraining. By making anticipatory DRL transparent and tunable, SIA lowers the barrier to proactive control in next-generation mobile networks.
\end{abstract}

\acresetall
% = = = = = = = = = = = = = = = = = =
\vspace*{-0.75ex}
\section{Introduction}
\label{sec:intro}
\vspace*{-0.75ex}
% = = = = = = = = = = = = = = = = = =
Next-generation mobile networks promise significant performance improvements but must cope with growing traffic demands and rapidly changing network conditions~\cite{tataria2021sixg, zhang2019vtm6g}. \ac{drl} is a promising approach for adaptive network control, with proven successes in resource allocation~\cite{mao2016resource, cai2023deep}, scheduling~\cite{mimo-agent, ye2017power}, and parameter optimization~\cite{ge-chroma-mobicom23, vannella2024learning}. \ac{drl} agents learn, through trial and error, policies that maximize cumulative rewards by weighting both the immediate and long-term outcomes of their actions~\cite{sutton2018reinforcement}.

A fundamental limitation, however, stems from how \ac{drl} agents interact with their environment. Mobile networks feature two distinct classes of \acp{kpi} with different relationships to agent actions~\cite{dietterich2000hierarchical,ng1999policy}. \textit{Controllable \acp{kpi}} respond directly to an agent's decisions. For example, scheduling a user directly affects their transmitted data, and activating base station antennas impacts power consumption. Agents naturally learn to predict how their actions influence these controllable \acp{kpi} using temporal difference learning~\cite{sutton1999between, sutton1988learning}. In contrast, \textit{exogenous \acp{kpi}} change regardless of the agent's actions~\cite{chen2018auto}. Factors like channel conditions shifting with user mobility or available bandwidth fluctuating due to external network congestion~\cite{pensieve} fall into this category. This leaves agents with a \textit{temporal myopia} when dealing with exogenous \acp{kpi}~\cite{worthy2012working}; they are blind to upcoming changes and can only react after they happen~\cite{arumugam2021deciding}.

\begin{figure}[t]
\centering%
\hspace*{-5pt}
\includegraphics[width=0.78\columnwidth,keepaspectratio]{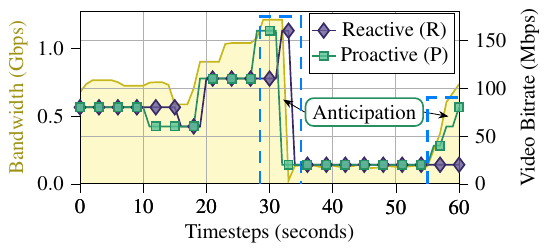}~%
\includegraphics[width=0.23\columnwidth,keepaspectratio]{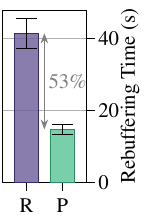}%
\vspace*{-1.5ex}%
\caption{Agents using future network bandwidth estimates achieve higher \ac{qoe} by acting proactively.}
\label{fig:intro-plot}%
\end{figure}

This reactive behavior leads to poor performance. In \ac{abr}, for instance, an agent learns how bitrate choices affect buffer levels (a controllable \ac{kpi}) but cannot anticipate sudden bandwidth drops (an exogenous \ac{kpi}). Such drops cause rebuffering~\cite{chen2023deeprophet}, which severely degrades user \ac{qoe}~\cite{k2024unveiling, bampis2017learning, krishnan2012video}. Recent approaches tackle this limitation by learning internal predictive models~\cite{fujimoto2023sale} or directly equipping agents with forecasts of future exogenous \acp{kpi}~\cite{chinchali2018cellular,lumos2023forc-tput8}. As Figure~\ref{fig:intro-plot} shows, a forecast-augmented agent can anticipate a bandwidth drop and proactively reduce its bitrate, reducing rebuffering time by $53$\%.

However, equipping agents with forecasts creates a new challenge. While performance improves, the lack of insight into agent behavior is a major barrier to real-world deployment~\cite{gunning2019darpa}, making it difficult to justify the added computational cost. This opacity raises critical questions: \textit{Do agents truly use forecasts to inform their decisions, or do they rely on current observations mostly?} \textit{When forecasts are used, how far ahead does an agent look, and does this horizon change with network dynamics?} Finally, \textit{how do forecasts alter an agent's strategy, for example, by causing it to sacrifice short-term gains for long-term benefits?} Understanding these trade-offs is essential to align agent behavior with network operator goals~\cite{clemm2022intent}.

These questions fall within the domain of \ac{xai}, a sub-field of \ac{ai} that develops techniques to interpret model decisions. However, existing methods in this field are designed specifically for reactive agents and cannot separate the influence of current observations from the impact of future predictions~\cite{ribeiro2016should, lundberg2017unified}. This critical diagnostic gap hinders the trust, debugging, and optimization of these forecast-augmented systems.

To close this gap, we present \toolname\ (Symbolic Interpretability for Anticipatory \ac{drl}), a framework designed to reveal how \ac{drl} agents exploit forecasts. Its primary goal is to transform opaque neural network decisions into human-interpretable explanations using \ac{sai} and per-\ac{kpi} knowledge graphs. 
Additionally, \toolname includes an optional Action Refinement module that uses forecasted values to improve an existing agent's performance, eliminating the need for costly retraining.
Our core contributions are:

\begin{itemize}[leftmargin=*]

    \item We design and implement \toolname, a novel interpretation framework for anticipatory \ac{drl} agents. By using symbolic abstractions and scalable, per-\ac{kpi} \acp{kg}, \toolname uniquely isolates the influence of current observations from future predictions in real time.

    \item We introduce the \ac{is}, a computationally efficient local explanation metric derived from our symbolic framework. The \ac{is} is the first metric to quantify and disentangle the influence of a \ac{kpi}'s current state from its predicted future trend.
    
    % \item We evaluate \toolname\ across three diverse use cases: \ac{abr}~\cite{pensieve} in the application layer, massive \ac{mimo} scheduling~\cite{mimo-agent} in the physical layer, and \ac{ran} slicing~\cite{polese2022colo} in the MAC layer.
    \item We demonstrate \toolname's practical impact across three diverse networking tasks: \ac{abr} streaming~\cite{pensieve}, massive \ac{mimo} scheduling~\cite{mimo-agent}, and \ac{ran} slicing~\cite{polese2022colo}. The insights from \toolname guide an \ac{abr} agent redesign that improves average bitrate by 9\%, while its Action Refinement module boosts the reward for a \ac{ran}-slicing agent by 25\% without retraining.
    
    % \item We demonstrate that \toolname's insights deliver concrete performance gains. A \toolname-guided redesign of \ac{abr} agent improves its average bitrate by 9\%, while the framework's Action Refinement module boosts the reward of \ac{ran}-slicing agent by 25\% without retraining.
    
    % \item We demonstrate \toolname's practical impact: its insights guide design modifications improving by 9\% the average bitrate in \ac{abr}, while its Action Refinement module provides 25\% reward increase for \ac{ran} slicing without costly retraining.
    % \item We demonstrate the practical impact of \toolname's insights, enabling significant improvements to existing agents either through design modification or action refinement—bypassing costly retraining and addressing a key deployment hurdle.
    % Our analysis uncovers sophisticated, previously hidden decision-making patterns. For instance, the streaming agent learns to select conservative bitrates to ensure smooth playback initially, then switches to aggressive quality maximization once the rebuffering risk is low. The \ac{mimo} scheduler exhibits similar temporal reasoning, deliberately accepting short-term interference for long-term throughput gains when forecasts indicate improving conditions. These insights directly lead to measurable gains; \toolname-guided design changes boost the \ac{abr} agent's average bitrate by $9$\%, while the Action Refinement module increases the \ac{ran} slicing agent's reward by $25$\% without retraining. 
\end{itemize}
Our code is publicly available at \url{https://github.com/RAINet-Lab/SIA}. 

% = = = = = = = = = = = = = = = = = =
\vspace*{-0.75ex}
\section{Background and Related Work}
\label{sec:background}
\vspace*{-0.75ex}
% = = = = = = = = = = = = = = = = = =

\subsection{Background on Anticipatory \ac{drl}}%: From Reactivity to Foresight}
\label{subsec:anticipatory_drl}
\vspace*{-0.5ex}

\ac{drl} agents learn optimal policies through dynamic interaction with an environment. At each timestep $t$, an agent in state $s_t \in \mathcal{S}$ selects action $a_t \in \mathcal{A}$ using its policy $\pi(a_t \mid s_t)$ and receives a reward $r_t$. The agent's goal is to maximize the cumulative reward $\mathbb{E}[\sum_{t=0}^{\infty} \gamma^t r_t]$, where $\gamma$ is a discount factor~\cite{sutton2018reinforcement}. The value function, $V^\pi(s) = \mathbb{E}[r_t + \gamma V^\pi(s_{t+1}) \mid s_t = s]$, provides long-term reasoning, helping agents anticipate how their actions affect \textit{controllable} \acp{kpi}. The controllable part of the next state, $s_{t+1}^c$, is a function of the agent's decision, $s_{t+1}^c = f(s_t, a_t)$, giving it direct influence over these outcomes.

This anticipation fails for \textit{exogenous} \acp{kpi}, where the transition to the exogenous part of the next state, $s_{t+1}^e$, is independent of the agent's actions ($s_{t+1}^e \perp a_t$). In these cases, \ac{drl} agents can only react to environmental changes. In video streaming, for example, an agent can learn to manage buffer levels (a controllable \ac{kpi}) through bitrate selection, but it cannot preempt disruptions from a sudden drop in network bandwidth (an exogenous \ac{kpi}).

To overcome this limitation, anticipatory \ac{drl} equips agents with forecasts of exogenous \acp{kpi}~\cite{morari1988model, moerland2023model}. A forecaster $g$ predicts future states $\hat{s}_{t+i}^e = g(s_{t-L:t}^e)$ for $i \in [1, h]$, where $L$ is the lookback window and $h$ is the prediction horizon. The agent then uses an augmented state $\overline{s}_t = [s_t, \hat{s}_{t+1}^e, \ldots, \hat{s}_{t+h}^e]$ to learn a proactive policy $\overline{\pi}: \overline{\mathcal{S}} \rightarrow \mathcal{A}$. For instance, LUMOS~\cite{lumos2023forc-tput8}, a decision-tree-based throughput predictor, showed that augmenting a model-predictive control (MPC) agent with its forecasts resulted in a 19.2\% \ac{qoe} improvement over 
baseline \ac{abr} algorithm.
Similarly, congestion forecasting has been shown to enable 14\% cellular capacity gains in load balancing~\cite{chinchali2018cellular}. Other techniques include learning predictive latent-space models~\cite{fujimoto2023sale} or redefining the state to include future predictions~\cite{singh2003learning,littman2001predictive}.

\vspace*{-1ex}
\subsection{Background on Symbolic AI}%: A Language for Interpretation}
\label{subsec:symbolic_ai}
\vspace*{-1ex}

\ac{sai} supports interpretable reasoning by encoding knowledge in explicit, human-readable structures~\cite{garcez2023neurosymbolic}. For anticipatory \ac{drl}, it can abstract raw numerical \acp{kpi} and actions into logical constructs that capture temporal relationships, such as state transitions and forecasted trends. This symbolic abstraction provides a clearer view of decision processes than analyzing the raw numerical \acp{kpi} the policy network operates on. \acf{fol}, a formal language for symbolic reasoning~\cite{russell2010artificial}, uses components like:

\vspace*{-1.5ex}%
\begin{tcolorbox}[colback=gray!15,colframe=gray!15,breakable,left=3pt,right=3pt]
$\bullet$ \textbf{Predicates}: Relations (e.g., $\mathit{allocate}$ or $\mathit{schedule}$)\par
$\bullet$ \textbf{Constants}: Domain entities (e.g., $\{Low, High\}$)\par
$\bullet$ \textbf{Variables}: Dynamic properties (e.g., $bandwidth$)\par
$\bullet$ \textbf{Quantifiers}: Scoping operators ($\forall, \exists$)\par
$\bullet$ \textbf{Connectives}: Logical operators ($\wedge, \vee, \neg$)
\end{tcolorbox}
\vspace*{-1.5ex}%

This formalism allows policies to be interpreted intuitively. Consider the rule: ``If latency exceeds 100\,ms \textit{and} packet loss exceeds 2\%, switch to the low-latency routing path.'' The corresponding \ac{fol} representation is:
\begin{align*}
\forall t \, [ &(\mathtt{highLatency}(t) \wedge \mathtt{highPacketLoss}(t)) \\
&\Rightarrow \mathtt{switchPath}(\texttt{lowLatency}) ]
\end{align*}
\begin{itemize}[leftmargin=*, topsep=0pt, itemsep=0pt]
\item $\mathtt{highLatency}(t)$: Latency $> 100$ms at $t$
\item $\mathtt{highPacketLoss}(t)$: Loss rate $> 2\%$ at $t$  
\item $\mathtt{switchPath}$: Action to change routing configuration
\end{itemize}

Raw measurements map directly to symbolic predicates, enabling both causal analysis and human-understandable rules~\cite{deraedt2020neurosymbolic, graves2016hybrid}. This transparency is essential for analyzing anticipatory \ac{drl} agents in networks.

\begin{figure*}[t]
\centering
\includegraphics[width=0.99\textwidth,keepaspectratio]{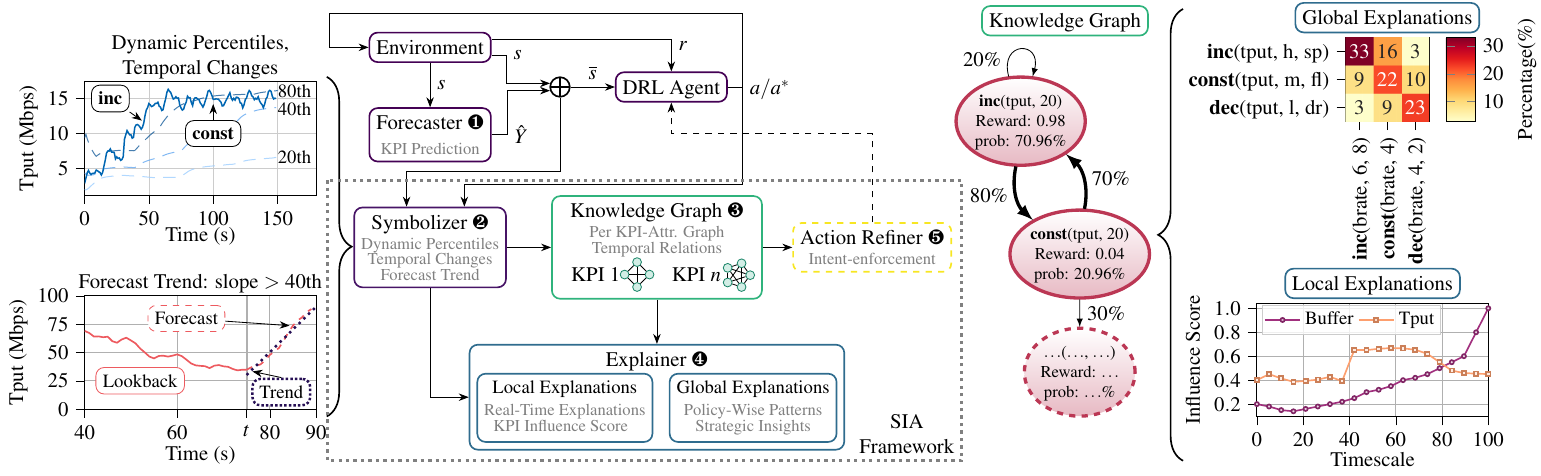}%
\vspace{-1ex}%
\caption{Architecture of \toolname, showing the core modules and information flow from raw \acp{kpi} to the explanations}%
\label{fig:framework-architecture}%
\vspace{-3ex}%
\end{figure*}

\vspace*{-1ex}
\subsection{Related Work}
\label{subsec:xai_gap}
\vspace*{-1ex}

\ac{xai} reveals how \ac{ai} models operate, helping build trust and enable debugging~\cite{gunning2019darpa}. However, most existing \ac{xai} approaches focus on reactive agents and therefore fail to capture the temporal complexity inherent in forecast-augmented \acp{drl}.

Post-hoc interpreters like \ac{lime}~\cite{ribeiro2016should} and \ac{shap}~\cite{lundberg2017unified} attribute importance to each of an agent's $k$ input features. Other approaches examine attention mechanisms~\cite{bibal2022attention} or analyze variable importance through gradient-based methods~\cite{ish2019interpreting}. However, these methods share common limitations: their utility in real-time network control is limited~\cite{milani2024explainable}.
Model-agnostic methods are computationally prohibitive, with methods like KernelSHAP having an exponential complexity in $k$, restricting them to offline analysis. While more efficient, model-specific variants exist, all versions share a more fundamental flaw: they treat all inputs as a static set, unable to distinguish the influence of current observations from future predictions. Even \ac{drl}-specific tools fall short. METIS~\cite{Metis} uses decision trees that ignore temporal constraints and requires slow retraining for updates. EXPLORA~\cite{fiandrino2023explora} builds massive state-action graphs that consume excessive memory and cannot handle new actions.

Other symbolic or programmatic approaches also fall short, as methods for rule extraction (e.g., NSRL~\cite{ma2021learning}), policy synthesis (e.g., PIRL~\cite{verma2018pirl}), or interpretation for reactive agents (e.g., SymbXRL~\cite{Dutt2505SymbXRL}) all lack the temporal awareness to disentangle the influence of current observations from future predictions. 
Finally, \ac{llm}-based methods~\cite{ameur2024leveraging,petroni2019lmkb}, despite having natural-language outputs, cannot meet the near-real-time demands (on the order of tens of milliseconds) of network-control loops and risk producing hallucinated explanations.

Two gaps therefore persist for agents that use forecasting. \textit{First}, existing tools lack the real-time performance needed for practical use in network-control operations. \textit{Second}, they are fundamentally unable to distinguish the influence of current observations from future predictions, a critical requirement for interpreting anticipatory agents. \toolname bridges these gaps by delivering efficient, temporally-grounded explanations for both proactive and reactive \ac{drl} agents. It uniquely reveals how forecasts reshape reactive decisions into proactive strategies through a flexible symbolic representation.

\vspace*{-0.75ex}
% = = = = = = = = = = = = = = = = = =
\section{The \toolname Framework}
\label{sec:sia-framework}%
% = = = = = = = = = = = = = = = = = =
\vspace*{-0.75ex}

Figure~\ref{fig:framework-architecture} illustrates \toolname's modular architecture, which is designed to operate alongside an anticipatory \ac{drl} agent. Raw \acp{kpi} pass through five modules: the Forecaster (\ding{182}) generates predictions; the Symbolizer (\ding{183}) converts them into symbolic representations; the \acf{kg} (\ding{184}) builds structured state-action relationships; the Explainer (\ding{185}) transforms these into human-readable insights; and finally, the optional Action Refiner (\ding{186}) uses these insights to improve agent decisions in real time.

\vspace*{-0.8ex}
\subsection{\toolname Core Modules}
\label{subsec:core-modules}
\vspace*{-0.8ex}

\subsubsection{Forecaster (\ding{182})}
\label{subsec:forecaster}

\toolname uses the existing forecasting component from a standard anticipatory \ac{drl} system (see~\S\ref{subsec:anticipatory_drl}). \toolname is agnostic to the forecast source; it uses the predictions ($\hat{Y}_t$) as provided, without modification. This module operates independently of other \toolname components, and can accommodate any forecasting approach (e.g., statistical, machine learning, or hybrid). As we demonstrate in \S\ref{sec:evaluation}, this modular design allows established forecasters to work effectively.%, giving operators flexibility without compatibility concerns.

\begin{algorithm}[t] 
\small
\caption{Symbolizer State Transformation Logic}
\label{alg:symbolizer}
\SetAlgoLined
\KwData{Current value $v_t$, previous value $v_{t-1}$, percentile sketch $P$, forecast series $F$, sensitivity $\theta$}
\KwResult{Symbolic state $s_{\text{sym}}$}

\emph{Change detection}\;
\eIf{$|v_t - v_{t-1}|/|v_{t-1}| > \theta$}{
    $predicate \gets
    \begin{cases}
        \mathtt{inc}   & \text{if } v_t > v_{t-1}\\
        \mathtt{dec}   & \text{otherwise}
    \end{cases}$\;
}{
    $predicate \gets \mathtt{const}$\;
}

\emph{Dynamic categorization}\;
$perc \gets \textsc{PercentileRank}(v_t,P)$\;
$category \gets \textsc{MapToBucket}(perc)$\;

\emph{Trend incorporation}\;
\eIf{$F \neq \emptyset$}{ 
    $slope \gets \textsc{LinReg}(F).\text{coef}$\;
    $s\_perc \gets \textsc{PercentileRank}(slope,\textit{SlopeHist})$\;
    $trend \gets \textsc{MapToTrend}(s\_perc)$\;
    \Return $\langle predicate, category, trend \rangle$\;
}{
    \Return $\langle predicate, category \rangle$\;
}
\end{algorithm}
% \vspace{-2ex} 

\subsubsection{Symbolizer (\ding{183})}
\label{subsec:symbolizer}

The Symbolizer converts raw  \acp{kpi} into symbolic representations using \ac{fol}, handling both current observations and forecasts. It consists of the three-stage logic detailed in Algorithm~\ref{alg:symbolizer}, namely:

\noindent\textbf{i) Change detection.}\; 
It identifies the direction of immediate changes between data points, as increasing (\lstinline!inc!), decreasing (\lstinline!dec!), or stable (\lstinline!const!) by applying a tunable sensitivity threshold, $\theta$ (default 5\%), to filter out insignificant noise.

\noindent\textbf{ii) Dynamic categorization.}\;
\ac{kpi} values are mapped to a configurable number of percentile-based categories. For example, a five-category bucket scheme would span 20 percentiles each (e.g., \lstinline!VeryLow! [0-20th], \lstinline!Low! [20-40th], etc.). This process adapts dynamically as network conditions evolve, with streaming quantile estimators (P$^2$) maintaining the required percentiles with $O(1)$~\cite{jain1985p2, greenwald2001space}.

\noindent\textbf{iii) Trend incorporation.}\; 
For forecasts with horizon $h{>}1$, the regression slope of the forecasted time-series is compared against a historical distribution of slopes and categorized into a configurable set of trends, such as \lstinline!Dropping! [<40th percentile], \lstinline!Fluctuating! [40-60th], or \lstinline!Spiking! [>60th].

\begin{lstlisting}[frame=lines,captionpos=t,caption={Example of state symbolization},label=lst:symbolizer-state]
# Throughput KPI symbolization
  current_tput  = 15.7  # Mbps, previous_tput = 12.3 # Mbps
# Change detection:
  change = (15.7 - 12.3) / 12.3 = 0.276 = 27.6%
  threshold = 5%
  27.6% > 5% => significant change detected
  15.7 > 12.3 => increasing
  predicate = inc
# Dynamic categorization:
  percentiles = [7.2, 10.5, 13.8, 18.2]  # P20,40,60,80
  15.7 Mbps in percentile rank => 72nd percentile
  Buckets: VeryLow[0-20], Low[20-40], Medium[40-60], 
             High[60-80], VeryHigh[80-100]
  72nd percentile => High category
# Trend incorporation:
  forecast = [14.9, 13.2, 11.8, 9.7]  # next 4 time steps
  linear regression slope = -1.68 Mbps/step
  slope percentile rank = 23rd percentile
  Buckets: Dropping[0-40], Fluctuating[40-60],
  Spiking[60-100]
  23rd percentile => Dropping trend
# Final symbolic state:
  => inc(tput, High, Dropping)
# Interpretation: "Throughput increased to High level but 
# is forecasted to be Dropping"
\end{lstlisting}

The symbolization process also extends to the agent's actions. The logic is configured based on the nature of the action space, allowing \toolname to adapt to different agents. For discrete, ordered actions like bitrate selection, it can represent the change relative to the previous action (e.g., \lstinline!dec(bitrate, 1200.0, 750.0)!). For categorical actions, such as selecting a scheduling policy, it can map to specific predicates (e.g., \lstinline!toPolicy(WF)!). This unified symbolic representation of both states and actions is a fundamental input for all subsequent modules in the \toolname framework.

Adapting \toolname to a new agent involves tuning three elements: i) the \textit{change-detection threshold}, $\theta$, to a \ac{kpi}'s volatility (e.g., a low $\theta$ for stable metrics like packet-loss versus a higher one for bursty signals like \ac{csi}); we found a consistent $\theta=5\,\%$ effective for our use cases; ii) the \textit{category scheme} to set the contextual resolution, where a given number of category buckets and their percentile boundaries are defined; and iii) the \textit{action representation}, which mirrors the agent's action space and can reuse standard \ac{kpi} logic or be customized for richer insights. This streamlined configuration allowed the same Symbolizer to operate across the \ac{abr}, \ac{mimo}, and \ac{ran}-slicing agents in our evaluations (see \S~\ref{sec:agent-scenario}).

\subsubsection{Knowledge Graph (\ding{184})}
\label{subsec:knowledge-graph}
The \acf{kg} module addresses the state-explosion problem common in existing \ac{xai} methods (see \S\ref{subsec:xai_gap}). While prior work often encodes the joint state space in a monolithic structure with exponential complexity, \toolname leverages \ac{kpi} independence by maintaining a separate directed, attributed graph for each \ac{kpi}. Nodes hold symbolic states (e.g., \lstinline!inc(tput,High,Dropping)!), edges represent agent actions, and attributes capture empirical action probabilities, reward estimates, and transition counts.

This design yields four key benefits:
(i) \textit{Bounded complexity}, as each graph contains at most 45 nodes ($3$ predicates $\times$ $5$ categories $\times$ $3$ trends);
(ii) \textit{Efficient graph updates}, with per-timestep time complexity $O(k)$. The per-\ac{kpi} factorization
{\setlength{\textfloatsep}{0pt}%
\begin{algorithm}[t] 
\small
\SetAlgoSkip{}%  <-- ADD THIS: removes algorithm2e's internal spacing
\caption{Forecast Aware Action Refinement}
\label{alg:action-refiner}
\SetAlgoLined
\KwData{Current state $s_t$; agent action $a_t$; forecasts $\hat{Y}$; knowledge graphs $\mathcal{KG}$; threshold $\tau$}
\KwResult{Refined action $a_{\text{refined}}$}

$s_{\text{current}} \gets \text{Symbolizer}(s_t)$\;
$s_{\text{future}} \gets \text{Symbolizer}(\hat{Y})$\;
$a_{\text{refined}} \gets a_t$ \tcp*{default: keep the agent action}
\For{each KPI $k$ with a forecast}{
    \If{edge $(s_{\text{current}}^k \rightarrow s_{\text{future}}^k) \in \mathcal{KG}_k$}{
        $a_{\text{best}}^k \gets \arg\max_a \bar{R}(a \mid s_{\text{current}}^k \!\rightarrow\! s_{\text{future}}^k)$\;
        $r_{\text{best}} \gets \bar{R}(a_{\text{best}}^k \mid s_{\text{current}}^k \!\rightarrow\! s_{\text{future}}^k)$\;
        $r_{\text{agent}} \gets \bar{R}(a_t \mid s_{\text{current}}^k)$ \textbf{if available, else} $0$\;
        \If{$r_{\text{best}} > r_{\text{agent}} + \tau$}{
            $a_{\text{refined}} \gets a_{\text{best}}^k$ \tcp*{override with the better action}
            \textbf{break}\;
        }
    }
}
\Return $a_{\text{refined}}$\;
\end{algorithm}}%
% \vspace{-2ex}%  <-- ADD THIS: additional negative space after the group
enables independent, parallelizable operations (see \S~\ref{subsec:performance-eval});
(iii) \textit{Causal querying}, for agent-strategy queries, such as determining the most likely actions when throughput is \lstinline!High! now but the forecasted trend is \lstinline!Dropping!;
and (iv) \textit{Efficient memory usage}, with space complexity $O(k,|S_{\text{sym}}|,|\mathcal{A}|)$, an exponential reduction from $O(|S|^k|\mathcal{A}|)$ in monolithic approaches, enabling real-time operation at scale.

\subsubsection{Explainer (\ding{185})}
\label{subsec:explainer}

The Explainer module converts the symbolic representations and \ac{kg} data into human-readable insights. It uses (1) the per-\ac{kpi} \acp{kg}; (2) the agent's current action; and (3) historical action distributions; to produce two complementary explanation types.

\textbf{Local explanations.} 
These quantify each \ac{kpi}'s influence on an individual decision using the \ac{is} (see \S\ref{subsec:local-explanations}). In contrast to perturbation-based analyses like \ac{shap} or \ac{lime}, which must generate and analyze many variations of each data point, our method efficiently reuses pre-computed \ac{kg} statistics. It runs in $O(k)$ time, returning an explanation in approximately $0.65$~ms per decision (see \S\ref{subsec:performance-eval}).

\textbf{Global explanations.} These reveal policy-level patterns. First, mutual information analysis between \acp{kpi} and actions pinpoints which metrics most strongly steer decisions. Second, action-focused policy graphs visualize state-action sequences, exposing biases and strategies that traditional feature-importance tools often miss.

\subsubsection{Action Refiner (\ding{186})}
\label{subsec:action-refinement}

The Action Refiner module augments a reactive agent with forecast awareness without retraining. As Algorithm~\ref{alg:action-refiner} shows, the module first symbolizes the current and forecasted states (lines 1–2). It then queries the per-\ac{kpi} \ac{kg} for each forecasted \ac{kpi} to find the historically best action, $a_{\text{best}}^k$, for the corresponding state transition (lines 4–6). If this proactive action’s expected reward exceeds the agent’s original choice by at least $\tau$, the module overrides the decision (lines 8–10). The computational complexity of this process is $O(k_f \times |\mathcal{A}|)$, where $k_f$ is the number of forecasted \acp{kpi} and $|\mathcal{A}|$ is the action-space size. We report the empirical latency measurements and a breakdown of per-component costs for this process in \S\ref{subsec:performance-eval}.

\vspace*{-1ex}
\subsection{\toolname's Explanations}
\label{subsec:explanation-methods}
\vspace*{-0.9ex}

\toolname's unified pipeline generates both local and global explanations from its symbolic representations, a key distinction from other \ac{xai} tools. Interpreters like \ac{lime} provide only local scores, while others like METIS and EXPLORA offer only global summaries. Although SHAP can produce both, it does so by aggregating numerous computationally expensive local scores. \toolname, in contrast, derives both explanations efficiently from its \ac{kg} structure, without altering the agent.

\noindent\textbf{Notation.} Let $\mathcal{K}$ be the set of all \acp{kpi}. For a given \ac{kpi} $k \in \mathcal{K}$, its symbolic state is $s_k \in \mathcal{S}_k$. The agent selects an action $a_t \in \mathcal{A}$ from the set of all possible actions at time $t$.

\subsubsection{Global Explanations}
\label{subsec:global-explanations}

Global explanations reveal the agent's overall policy through two complementary techniques.

\textbf{Mutual-information analysis} quantifies the statistical dependence between each symbolic \ac{kpi} and the action distribution:
\begin{equation}
MI(k; a)=\sum_{s_k \in \mathcal{S}_k}\sum_{a \in \mathcal{A}} p(s_k,a)
    \log\frac{p(s_k,a)}{p(s_k)\,p(a)},\quad k \in \mathcal{K}.
\label{eq:mutual_information}
\end{equation}

Here, $p(s_k, a)$ is the joint probability, while $p(s_k)$ and $p(a)$ are the marginals. Ideal for symbolic data, this metric captures nonlinear relationships without distributional assumptions or discretization and has low $O(|\mathcal{S}_k|\,|\mathcal{A}|)$ complexity.

\textbf{Action-focused policy graphs} visualize decision-making logic by representing actions as nodes and state transitions as directed edges. As shown in \S\ref{subsubsec:policy-analysis}, these graphs expose hidden strategies and biases by annotating transitions with metrics like frequency and average reward.

\subsubsection{Local Explanations: The Influence Score}
\label{subsec:local-explanations}

For real-time decision analysis, we introduce the \acf{is} to quantify each \ac{kpi}'s contribution to a specific action. This score is derived from the per-\ac{kpi} \ac{kg} in four steps:

\begin{enumerate}[leftmargin=*, wide=0\parindent,label=\Roman*)]

    \item \textbf{Extract Conditional Probabilities:} For each \ac{kpi} $k$, compute its conditional action distribution, $P_k(a|s_k) = \frac{\text{count}(a, s_k)}{\text{count}(s_k)}$, where $\text{count}(a, s_k)$ is the historical count derived from node and edge annotations in the \ac{kpi}'s \ac{kg}.

    \item \textbf{Establish a Baseline:} Compute a baseline action distribution, $P_\emptyset(a) = \frac{1}{|\mathcal{K}|} \sum_{k \in \mathcal{K}} P_k(a|s_k)$, which serves as the average action probability across all \acp{kpi}.

    \item \textbf{Calculate Information Contribution:} Measure the influence of a \ac{kpi} using its \ac{kl}-divergence from the baseline ($D_{\textrm{KL}}(P_k || P_\emptyset)$), which quantifies how much the \ac{kpi}'s state reduces uncertainty about the agent's action.

    \item \textbf{Apply Alignment Weighting:} Weight the information contribution by an alignment function, $\delta(a_t, a_k^*)$, which filters the influence based on how well the agent's chosen action $a_t$ aligns with the \ac{kpi}'s most likely action, $a_k^* = \arg\max_{a} P_k(a|s_k)$.

\end{enumerate}

Conceptually, a large divergence means that the \ac{kpi} provides a distinctive signal relative to average behavior. The complete formulation combines these steps:
\begin{equation}
    \textrm{IS}_k = D_{\textrm{KL}}(P_k || P_\emptyset) \times \delta(a_t, a_k^*).
\label{eq:influence_score}
\end{equation}

The alignment function $\delta$ adapts the \ac{is} to action types. For continuous actions (e.g., power allocation, bitrate selection), we use $\delta_{\text{decay}}(a_t, a_k^*) = \exp(-d(a_t, a_k^*)^2/2\sigma^2)$, where $d(\cdot,\cdot)$ measures action distance and $\sigma$ controls sensitivity. This approach achieves $O(k)$ complexity for bounded action spaces, enabling real-time explanations that capture temporal dependencies, a capability absent in static methods like \ac{shap} or \ac{lime}.

% = = = = = = = = = = = = = = = = = =
\vspace*{-0.75ex}
\section{Use Cases and Experimental Setup}
\label{sec:agent-scenario}
\vspace*{-0.75ex}
% = = = = = = = = = = = = = = = = = =

To evaluate \toolname, we test it across three distinct use cases spanning different network layers: \ac{abr}, Massive \ac{mimo} scheduling, and \ac{ran} slicing. This diversity demonstrates \toolname's flexibility. In this section, we detail each use case, including the \ac{drl} agent's design, state, reward function, and the forecasting model chosen for the task. While the agents are heterogeneous, \toolname unifies them by converting their raw \acp{kpi} into a consistent symbolic representation based on \ac{fol}.

Our evaluation is guided by three research questions (RQs):
\begin{enumerate}[label=$\bullet$ \textit{RQ$_{\arabic*}$}:, ref=\textit{RQ$_{\arabic*}$}, wide=0\parindent, listparindent=0pt, align=left]
    \item \label{rq1} How does \toolname explain a forecast-aware agent's policy compared to a reactive one, and how do these explanations improve agent performance?
    \item \label{rq2} How does \toolname's \ac{is} compare against established interpreters like SHAP and LIME?
    \item \label{rq3} How effective is \toolname's Action Refinement module at boosting performance without retraining the agent?
\end{enumerate}

\begin{figure}[t]
\centering
\includegraphics[width=1\columnwidth,keepaspectratio]{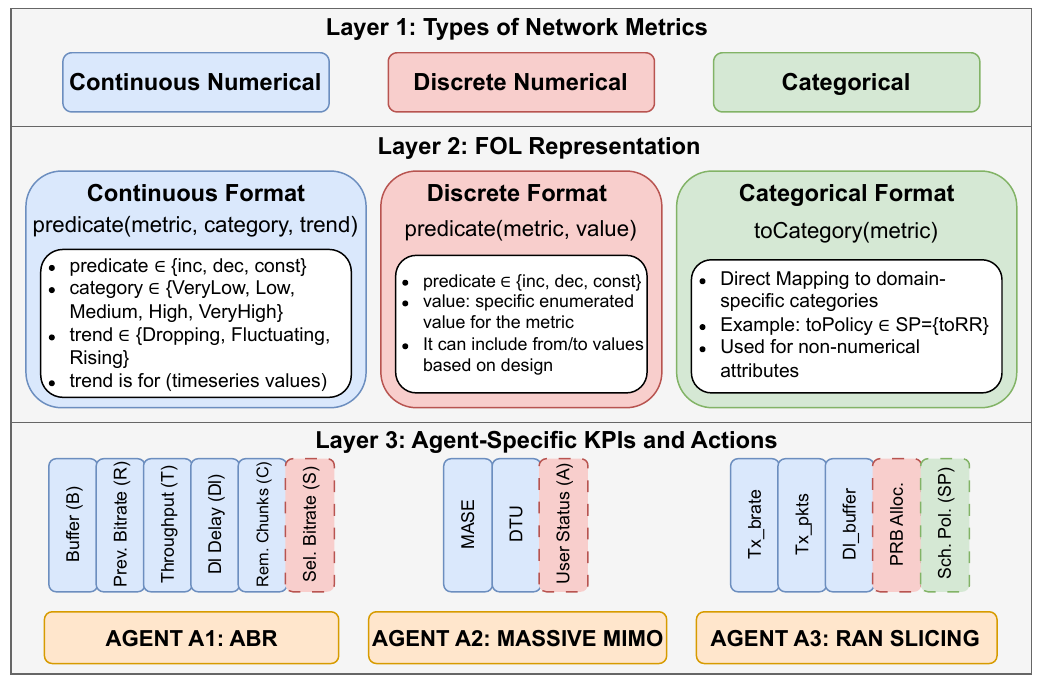}%
\vspace*{-1.5ex}%
\caption{Symbolic \ac{fol} representations for the \acp{kpi} of our evaluation agents.}
\label{fig:fol-conv}%
\vspace*{-1.5ex}
\end{figure}

Figure~\ref{fig:fol-conv} illustrates the general \ac{fol} representations used across all agents. Continuous \acp{kpi} are mapped to \lstinline!predicate(metric, category, trend)!, discrete values to \lstinline!predicate(metric, value)!, and categorical variables to \lstinline!toCategory(metric)!.

\vspace*{-1ex}
\subsection{A1: ABR for Video Streaming}
\label{agent-a1}
\vspace*{-1ex}

For our first use case, we adapt the Pensieve \ac{drl} agent~\cite{pensieve} to an \ac{abr} scenario that aims to maximize video quality while minimizing stalling. The agent's state includes buffer levels, the last chosen bitrate, download delay and throughput (Thput) histories (e.g., symbolically as \lstinline!inc(tput, High, Dropping)!), and the number of chunks remaining. The agent's action is to select the next bitrate from a discrete set, while its reward function balances the competing goals of high bitrate, low rebuffering, and smooth quality transitions:
\begin{equation}
\label{ABR-REward}
R_t = \underbrace{\sum_{n=1}^{N} q(R_n)}_\textrm{Bitrate Utility} - \underbrace{\mu \sum_{n=1}^{N} T_n}_\textrm{Rebuffering} - \underbrace{\sum_{n=1}^{N-1} \lvert q(R_{n+1}) - q(R_n) \rvert}_\textrm{Quality Variation(Smoothness)}.
\end{equation}

To give the agent foresight, we equip it with forecasts of exogenous \acp{kpi}, selecting the optimal model for each task. For the \AoneP agent, which performs \textit{univariate} forecasting of bandwidth (Bwd), we employ PatchTST~\cite{patch-tst}, a \ac{sota} Transformer model recognized for its superior performance in long-horizon univariate forecasting. In contrast, the \AonePSIA agent is redesigned using insights from \toolname (see Section~\ref{subsubsec:mi-analysis}) and requires a simultaneous, \textit{multivariate} forecasting of Bwd, Thput, and delay. For this, we use a lightweight \ac{mlp} paired with Reversible Instance Normalization (RevIN)~\cite{revin}. This choice is data-driven: our evaluation shows the MLP-RevIN model achieves 20\% \ac{mape}, significantly outperforming alternatives from literature like Lumos (80\% \ac{mape})~\cite{lumos2023forc-tput8} and Xatu (30\% \ac{mape})~\cite{xatu}. All agents are tested on datasets from~\cite{Lumos5G-dataset,Norway-abrdataset}; Table~\ref{tab:agent-configs} summarizes their configurations.

\vspace*{-1ex}
\subsection{A2: User Scheduling in Massive MIMO}
\label{agent-a2}
\vspace*{-1ex}

In this use case, we use the Massive \ac{mimo} scheduler from~\cite{mimo-agent}. The agent's action is to decide for each of the seven users whether to schedule them or not. To interpret the policy's effect on interference, \toolname abstracts these collective actions into a higher-level symbolic representation, \texttt{Alloc(User Group, Percentage)}, which captures the percentage of resources given to each user group. The agent's state includes per-user \acp{kpi} like \ac{mase}, \ac{dtu}, and their assigned \ac{ug}. Agent's reward signal balances system throughput and \ac{jfi}~\cite{jain1984quantitative}, as
\begin{equation}
\label{MIMO-REward}
R_t = \underbrace{\beta \gamma_t^{\text{total}}}_\textrm{System Data Transmission} + \underbrace{(1 - \beta) \text{JFI}_t}_\textrm{Data Fairness}, \quad \beta = 0.5.
\end{equation}

To predict the exogenous \ac{mase} \ac{kpi} for the forecast-augmented agent (\AtwoP), we again use PatchTST~\cite{patch-tst}, as this is another univariate forecasting task. By using these forecasts, the anticipatory agent learns an effective policy 24 episodes faster than the reactive baseline (\AtwoR).

\vspace*{-1ex}
\subsection{A3: \ac{ran} Slicing and Policy Scheduling}
\label{agent-a3}
\vspace*{-1ex}
Our third use case, from~\cite{polese2022colo}, involves joint \ac{ran} slicing and scheduling for three traffic slices, running as an xApp on the O-RAN near-RT RIC and emulated in Colosseum~\cite{polese2024colosseum}. State \acp{kpi} include: \txbrate (e.g., \lstinline!inc(tx_brate, High)!), \txpkts, and \dlbuff, while actions control \ac{prb} allocation and scheduling policies for each slice. We chose this complex agent to test \toolname's action refinement ability, as retraining is impractical due to the large action space. To forecast the exogenous \acp{kpi} (\txbrate and \dlbuff), we use a multivariate \ac{mlp} architecture paired with RevIN.

\begin{table}[t]
\centering
\setlength{\tabcolsep}{4pt}
\caption{Agent Configurations. $h$ denotes the forecast horizon}
\label{tab:agent-configs}
\vspace{-1.5ex}
\begin{tabularx}{\columnwidth}{@{} ll >{\raggedright\arraybackslash}X l c @{}}
\toprule
\textbf{Agent} & \textbf{Config} & \textbf{Forecasted \acp{kpi}} & \textbf{Forecast Model} & \textbf{$\bm{h}$} \\
\midrule
\multicolumn{5}{@{}l}{\textit{A1: ABR Streaming~\cite{pensieve}}} \\
\AoneR      & Reactive       & None                   & N/A       & 0 \\
\AoneP      & Proactive      & Bwd                    & PatchTST  & 4 \\
\AonePSIA   & \toolname-guided design     & Bwd, Thput, Delay      & MLP-RevIN & 4 \\
\midrule
\multicolumn{5}{@{}l}{\textit{A2: MIMO Scheduling~\cite{mimo-agent}}} \\
\AtwoR      & Reactive       & None                   & N/A       & 0 \\
\AtwoP      & Proactive      & MASE                   & PatchTST  & 4 \\
\AtwoRSIA   & Action-Refined & MASE                   & PatchTST  & 4 \\
\midrule
\multicolumn{5}{@{}l}{\textit{A3: RAN Slicing~\cite{polese2022colo}}} \\
\AthreeR    & Reactive       & None                   & N/A       & 0 \\
\AthreeRSIA & Action-Refined & tx\_brate, dl\_buffer    & MLP-RevIN & 4 \\
\bottomrule
\end{tabularx}
\vspace{-1.5ex}
\end{table}

The Symbolizer's parameters were chosen to balance detail and stability. 
The change-detection threshold, $\theta$, is configurable for each \ac{kpi} to handle different numerical scales; we found $\theta=5\%$ effective for our use cases. 
The number of percentile buckets is also configurable and chosen for desired granularity: five for primary \acp{kpi} and three for forecasted trends.

% #######################################################################################
% Evaluation Results
% #######################################################################################
% = = = = = = = = = = = = = = = = = =

\vspace*{-0.75ex}
\section{Evaluation Results}
\label{sec:evaluation}
\vspace*{-0.75ex}
% = = = = = = = = = = = = = = = = = =

\begin{figure*}[t]
\begin{minipage}{0.5\textwidth}%
\centering
\includegraphics[width=0.95\columnwidth,keepaspectratio]{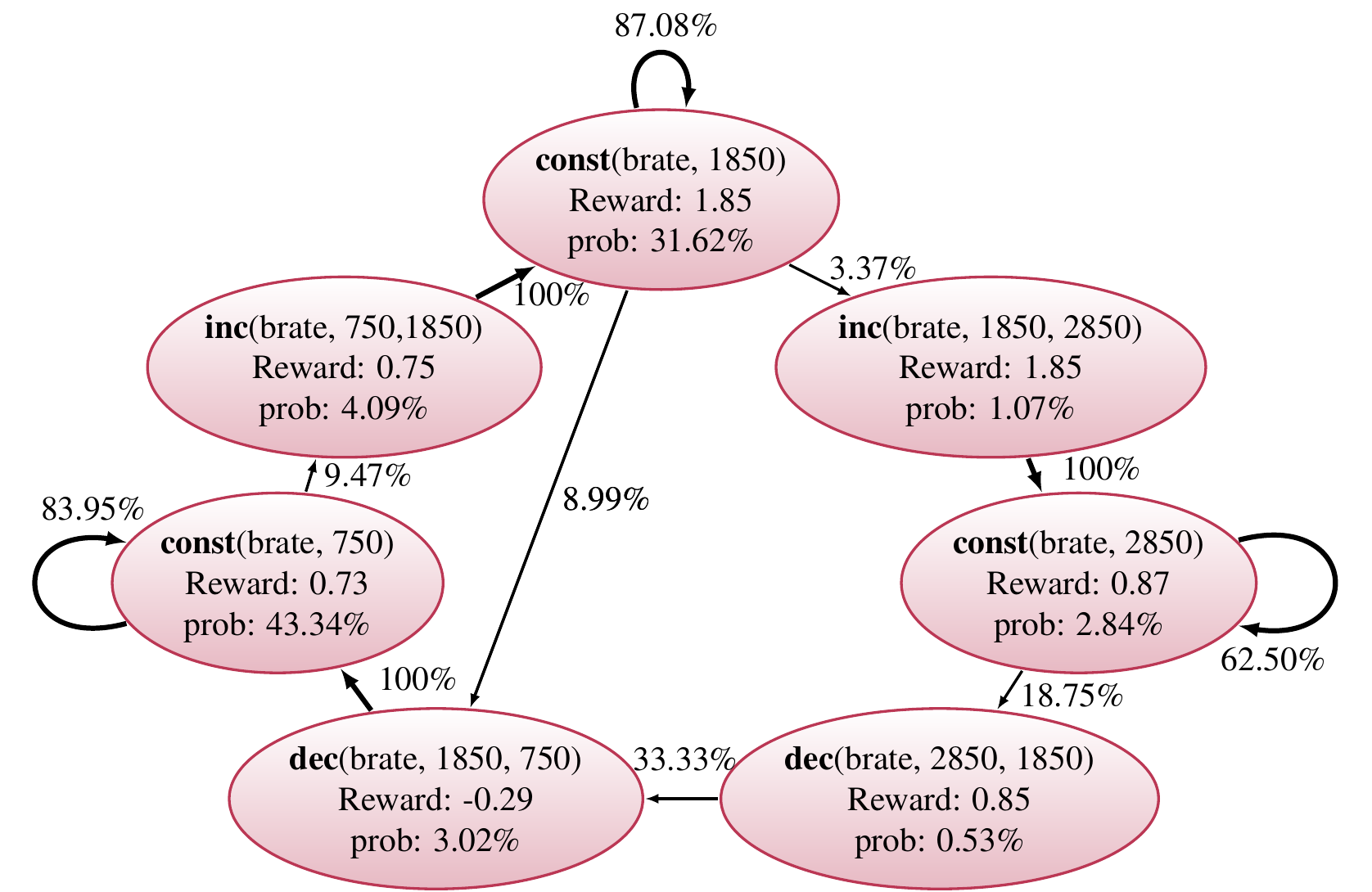}%
\caption{Policy graph of the reactive agent (\AoneR).}
\label{fig:pol-graph-pensieve-vanilla}
\end{minipage}%
\begin{minipage}{0.5\textwidth}%
\centering
\includegraphics[width=0.95\columnwidth,keepaspectratio]{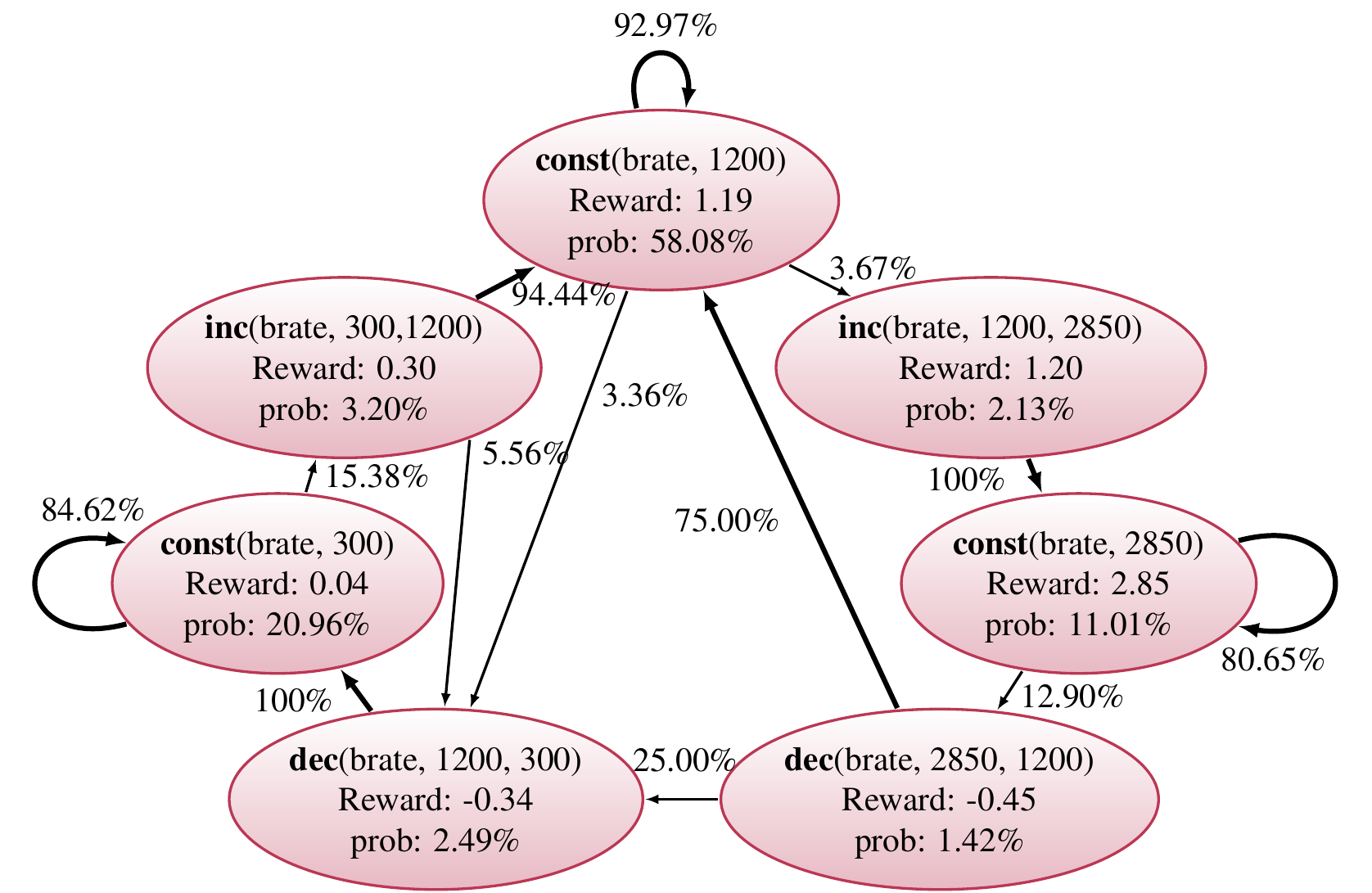}%
\caption{Policy graph of the proactive agent (\AoneP).}
\label{fig:pol-graph-pensieve-forecaster}
\end{minipage}
\vspace*{-4ex}%
\end{figure*}

This section evaluates \toolname using the agent configurations from \S\ref{sec:agent-scenario}. We use agents A1 and A2 to address \ref{rq1} and \ref{rq2}, and agents A2 and A3 for \ref{rq3}. We compare \toolname's explanations to \ac{sota} interpreters and demonstrate how it improves agent performance without retraining.

\vspace*{-1ex}
\subsection{Analysis of Global Explanations}
\label{subsec:results-global}
\vspace*{-1ex}

To address \textit{\ref{rq1}}, we demonstrate how \toolname's global explanations reveal anticipatory policies, uncover design flaws, and guide performance improvements using policy graphs and \ac{mi} analysis.

% \vspace*{-1ex}
\subsubsection{Agent Policy Analysis}
\label{subsubsec:policy-analysis}

\toolname's policy graphs visualize an agent's strategy by mapping symbolic actions to nodes and transitions to edges. Figures~\ref{fig:pol-graph-pensieve-vanilla} and~\ref{fig:pol-graph-pensieve-forecaster} compare the policies of the reactive (\AoneR) and proactive (\AoneP) agents in the \ac{abr} task~\cite{Norway-abrdataset}. The graphs reveal key behavioral differences:
\begin{itemize}[leftmargin=*]
    \item \AoneP maintains a higher steady-state quality, spending over $58$\% of its time at $1200$~kbps, while \AoneR defaults to a lower $750$~kbps baseline for $43$\% of its time.

    \item \AoneR uses incremental transitions ($750 \rightarrow 1850 \rightarrow 2850$~kbps) before retreating, whereas \AoneP makes decisive, forecast-guided shifts. This stability is evident in its stronger self-loops (e.g., an $80.65\%$ probability of staying at $2850$~kbps versus \AoneR's $62.5\%$).

    \item \AoneP leverages network forecasts to sustain the maximum bitrate for extended periods, spending nearly four times longer at peak quality than \AoneR ($11.01\%$ vs. $2.84\%$).

    \item \AoneP employs different recovery paths, such as dropping to $300$~kbps vs $750$~kbps for \AoneR to rebuild its buffer; this forward-looking strategy is missing in the reactive \AoneR.
    
\end{itemize}

These proactive strategies yield measurable gains: Table~\ref{tab:agent-a1-results} shows \AoneP achieves a $1.7$\% higher reward and a $1.8$\% higher bitrate. Unlike traditional interpreters like Metis~\cite{Metis}, which produce large decision trees ($3200$--$5000$ nodes) with long retraining cycles ($\sim30$~min), \toolname's bounded policy graphs offer an interpretable, real-time visualization of the agent's temporal action policies that are hidden by previous methods.

\begin{table}[t]
    \centering
    \caption{Performance of proactive agent (\AoneP) and \toolname guided proactive (\AonePSIA) compared to the reactive baseline (\AoneR).}
    \label{tab:agent-a1-results}
    \vspace{-2ex}
    \resizebox{0.95\columnwidth}{!}
    {
        \begin{tabular}{lccc}
            \toprule
            \textbf{Agent} & \textbf{Reward Increase} & \textbf{Bitrate Increase} & \textbf{P-value} \\
            \midrule
            \AoneP & +$1.7$\% & +$1.8$\% & < 0.01 \\
            \AonePSIA & +$3.4$\% & +$9.0$\% & < 0.01 \\
            \bottomrule
        \end{tabular}%
    }
    \vspace{-1.5ex}
\end{table}

\begin{figure}[t]
\centering
\includegraphics[width=0.95\columnwidth,keepaspectratio]{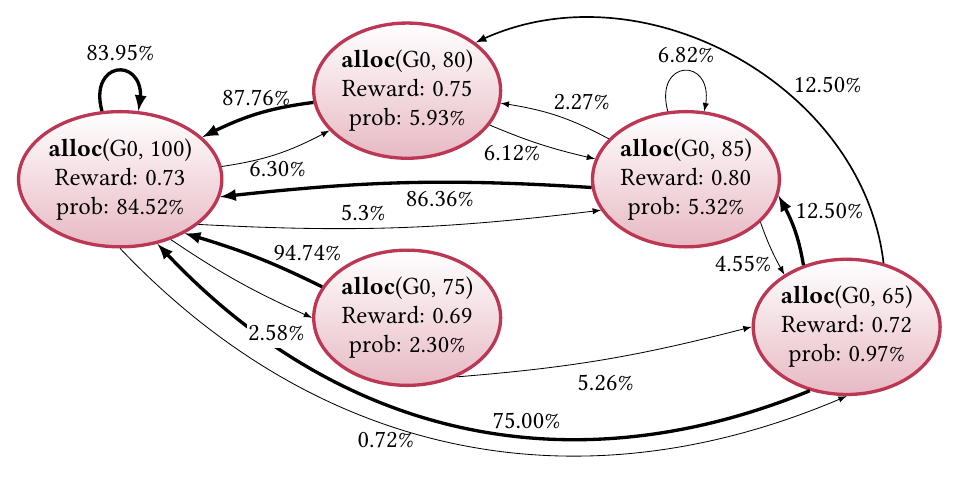}%
\vspace{-1ex}%
\caption{Policy graph of the reactive agent (\AtwoR) for Group 0 users.}%
\label{fig:policy-graph-mimo}%
\vspace{-1.5ex}%
\end{figure}

\subsubsection{Revealing Design Biases}
\label{subsubsec:design-bias}

\toolname's policy graphs can also expose subtle flaws in an agent's reward design. Figure~\ref{fig:policy-graph-mimo} shows the policy for the reactive \ac{mimo} scheduling agent (\AtwoR), which, by design, should aim to schedule users from Group 0 (group of users with the best channel quality).

The graph reveals a counterintuitive behavior: the agent's most probable action, allocating $100$\% of resources to Group 0, yields a lower reward ($0.73$) than a partial allocation ($0.80$ for an $85$\% allocation). This result is unexpected, as scheduling other groups should introduce interference. The root cause lies in the reward function (Eq.~\eqref{MIMO-REward}): the \ac{jfi} is based on cumulative \ac{dtu} and quickly saturates near $1.0$. As these values grow large, the index becomes insensitive to momentary allocation differences, causing the agent to prioritize short-term throughput at the cost of system-wide interference.

This apparent contradiction (why the agent favors an action with lower reward) is explained by opportunity. Multiple user groups are only present in $17$\% of timesteps. The policy flaw is thus revealed during these critical steps, where the agent diverts resources from Group 0 in $72$\% of cases, causing interference. Traditional feature-importance methods would miss this policy-level flaw, which \toolname exposes by showing how a faulty reward design can lead to a counter-productive policy.

\begin{figure*}[t]
\centering%
\includegraphics[width=0.9\textwidth,keepaspectratio]{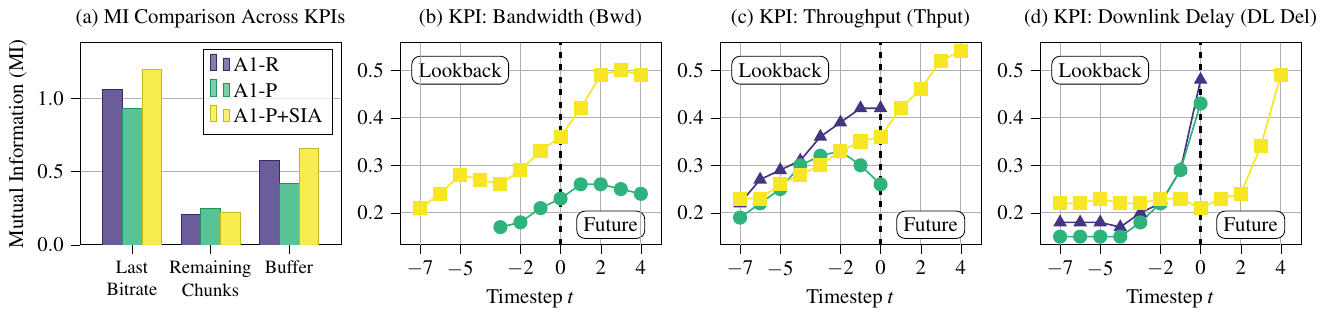}%
\vspace*{-2ex}%
\caption{\ac{mi} analysis between input \acp{kpi} and agent actions for the reactive (\AoneR), proactive (\AoneP), and \toolname-guided (\AonePSIA) agents.}
\label{fig:mi-analysis}%
\vspace*{-4ex}%
\end{figure*}

\subsubsection{Quantifying Forecast Impact Through \ac{mi}}
\label{subsubsec:mi-analysis}

We use \ac{mi} analysis to reveal a critical design flaw in the proactive agent (\AoneP) arising from naively integrating forecasts with inconsistent temporal structures. Figure~\ref{fig:mi-analysis} shows the \ac{mi} values for our three \ac{abr} agent variants.

The analysis of static \acp{kpi} shows consistent influence across all agents (Fig.~\ref{fig:mi-analysis}a), but a critical misalignment is exposed when examining the temporal \acp{kpi}. The issue is that the agent receives correlated inputs with inconsistent time horizons. For agent \AoneP, both bandwidth and throughput are fed as vectors of 8 timesteps. However, the bandwidth input is structured around its forecast, containing 3 past, 1 current, and 4 future values. In contrast, the throughput input retains its original reactive structure from agent \AoneR, containing only 7 past values and 1 current value, with no forecast. Figure~\ref{fig:mi-analysis}(b-d) shows the consequence: the agent learns from bandwidth that the strongest predictive signal comes from the values near its horizon ($t=1$ and $t=2$). It then incorrectly applies this heuristic to the throughput data, learning to focus on timesteps just before the horizon. Since the throughput horizon is at $t=0$, this causes the \ac{mi} peak to incorrectly shift to $t=-2$.

This suggests \textit{the agent learns relative temporal patterns rather than understanding time inherently}. Guided by this insight, we designed \AonePSIA with a \textit{uniform} temporal structure (7 past, 1 current, and 4 future values) across all correlated \acp{kpi}. This \toolname-guided redesign corrected the misalignment and, as shown in Table~\ref{tab:agent-a1-results}, produced substantial gains, achieving a $9$\% higher bitrate and a $3.4$\% higher overall reward than the baseline. This uncovers a key principle for anticipatory \ac{drl}: \textit{correlated temporal inputs must be presented with consistent formatting to prevent policy misinterpretation}.

\vspace*{-1ex}
\subsection{Analyses of Local Explanations}
\label{subsec:results-local}
\vspace*{-1ex}

To answer \textit{\ref{rq2}}, this section demonstrates \toolname's ability to generate real-time, temporally-aware local explanations. We show how the \ac{is} enables operational monitoring and reveals strategic adaptations that are invisible to interpreters like \ac{shap} and \ac{lime}.

\subsubsection{Real-Time Monitoring with Influence Scores}
\label{subsubsec:realtime-influence}

\begin{figure}[t]
\centering
\includegraphics[width=0.95\columnwidth,keepaspectratio]{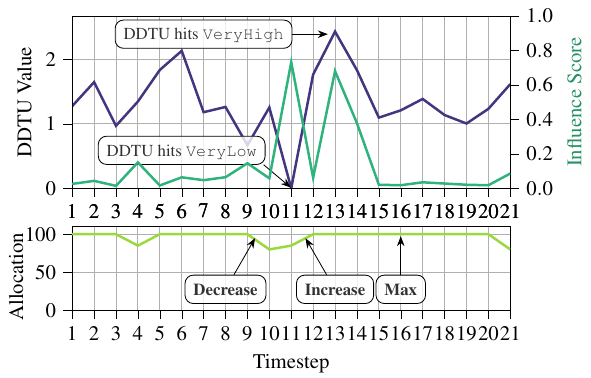}%
\vspace*{-3ex}%
\caption{\ac{is} for Agent \AtwoR's Decision in Group 0 and \ac{ddtu}'s Effect}%
\label{fig:example-ddtu-agent-decision}%
\vspace*{-1ex}%
\end{figure}

Figure~\ref{fig:example-ddtu-agent-decision} demonstrates \toolname's real-time monitoring of the reactive \ac{mimo} agent, \AtwoR. The plots track the \ac{ddtu} \ac{kpi}, its corresponding \ac{is}, and the agent's resource allocation for Group 0, revealing a clear causal chain. For instance, after an allocation drop at timestep 9 that causes the \ac{ddtu} to fall, its \ac{is} spikes to nearly 0.8 upon entering a \lstinline!VeryLow! category at timestep 11. This drop flags the \ac{ddtu} as the dominant factor, prompting an immediate increase in resource allocation for Group 0. This causal analysis, explaining both when and why an agent reacts, is generated in just $0.65$\,ms mean latency. In contrast, obtaining a single explanation from \ac{shap} or \ac{lime} on the same task takes $141$\,ms and $159$\,ms, respectively (see \S\ref{subsec:performance-eval}), enabling true real-time observability.

\subsubsection{Revealing Strategic Adaptation Over Time}
\label{subsubsec:temporal-influence-evolution}

\begin{figure}[t]
\centering
\includegraphics[width=0.95\columnwidth,keepaspectratio]{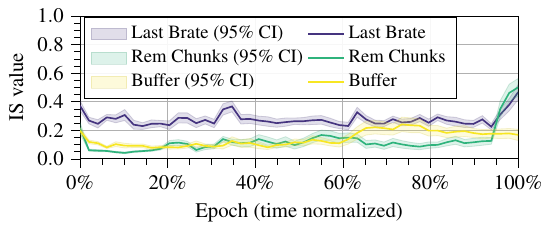}
\vspace*{-2ex}%
\caption{\ac{is} for agent \AoneR, showing \acp{kpi} importance}
\label{fig:influence-score}%
\vspace{-2ex}%
\end{figure}

Beyond explaining single decisions, the \ac{is} can reveal how an agent's strategy dynamically evolves over an entire episode. Figure~\ref{fig:influence-score} tracks the \ac{is} (y-axis) over the normalized episode duration (x-axis) for three key static \acp{kpi} influencing the reactive agent's (\AoneR) streaming strategy on the \acs{5g} dataset~\cite{Lumos5G-dataset}.

The analysis reveals a sophisticated, multi-faceted policy. In the initial phase (0--$80$\%), the agent prioritizes stable playback, evidenced by the dominance of the Last Bitrate \ac{kpi} ($\text{\ac{is}} \approx 0.3$), reflecting the smoothness term in the reward function~\eqref{ABR-REward}. Concurrently, the Buffer influence peaks at startup and again near the $60$\% mark, showing an intermittent focus on refilling buffer to prevent stalls, reflecting the rebuffering term~\eqref{ABR-REward}.

A dramatic shift occurs in the final quality maximization phase ($80$--$100$\%), where the influence of Remaining Chunks triples to become the dominant \ac{kpi} ($\text{\ac{is}} \approx 0.45$). With few chunks left, the rebuffering penalty risk diminishes, allowing the agent to aggressively pursue a higher bitrate. This adaptation suggests how network operators can align resources with the agent's policy: provide stable bandwidth initially and then offer bursty, high-throughput resources toward the episode's end.

\subsubsection{Distinguishing Temporal from Static Features}
\label{subsubsec:comparative-analysis}

\begin{figure}[t]
\centering
\includegraphics[width=0.95\columnwidth,keepaspectratio]{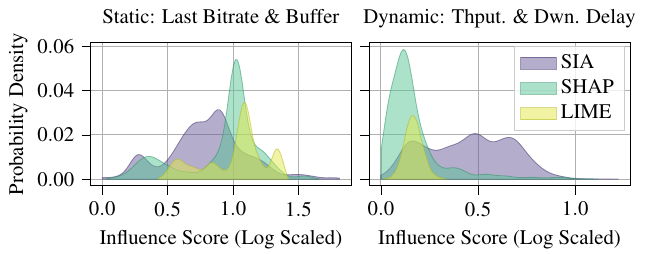}%
\vspace*{-2ex}
\caption{Comparing \ac{is} of different interpreters}%
\label{fig:prob-distributions}%
\vspace*{-1ex}%
\end{figure}

Figure~\ref{fig:prob-distributions} highlights \toolname's unique temporal awareness compared to \ac{shap} and \ac{lime}. While all methods produce similar importance distributions for static features (left panel), a clear distinction emerges for temporal ones (right panel). Here, \toolname's \ac{is} distribution is distinctly bimodal as a direct result of its symbolic encoding and the mechanics of its formula~\eqref{eq:influence_score}.

This bimodality arises because the symbolic state $s_k$ for a temporal \ac{kpi} is a composite of category and trend components. The \ac{is}, driven by the $D_{\textrm{KL}}$ term, yields a high score in two different scenarios. The first peak represents decisions driven by the \ac{kpi}'s current \textit{category} (e.g., throughput is \lstinline!VeryHigh!), while the second peak reflects decisions driven by its future \textit{trend} (e.g., throughput is \lstinline!Dropping!), which can contradict the current category.

For instance, \toolname can differentiate whether a bitrate reduction stems from a \lstinline!Low! current throughput (a category-driven decision) or from a \lstinline!Dropping! trend despite a \lstinline!High! current throughput (a trend-driven decision). In contrast, \ac{shap} and \ac{lime} average the importance across all temporal data points, blending these distinct contexts into a single, less informative peak. This ability to disentangle the influence of current state versus future predictions is a critical capability for explaining anticipatory agents, directly addressing \textit{\ref{rq2}}.

\vspace*{-1ex}
\subsection{Performance and Scalability Analysis}
\label{subsec:performance-eval}
\vspace*{-1ex}

\toolname is designed for real-time operation. We report worst-case performance, measured on our most complex agent (\AoneP), which has the highest number of input \acp{kpi}. The core pipeline for a local explanation, including symbolizing inputs, updating the \ac{kg}, and calculating the \ac{is}, completes in a mean time of just $0.65$\,ms. Table~\ref{tab:performance} details the sub-millisecond latency of each component. This makes the full process over $200\times$ faster than traditional interpreters like \ac{shap} ($141$\,ms) and \ac{lime} ($159$\,ms) on the same task.

This efficiency stems from \toolname's architecture. Because its complexity scales linearly ($O(k)$) with the number of \acp{kpi}, unlike the exponential complexity of competing methods, its performance is predictable. These worst-case latencies ensure \toolname operates comfortably within the one-second control loop of the O-RAN near-RT RIC constraints.

\begin{table}[t]
\centering
\caption{Latency per \toolname component and timestep of the \AoneP agent.}%
\label{tab:performance}%
\vspace*{-1ex}%
\resizebox{0.8\columnwidth}{!}{%
\begin{tabular}{lcc}
\toprule
\textbf{Component} & \textbf{Mean Latency} & \textbf{Std. Dev.} \\
\midrule
Symbolizer & $0.099$~ms & $0.009$~ms \\
Knowledge Graph Update & $0.265$~ms & $0.055$~ms \\
Influence Score & $0.280$~ms & $0.030$~ms \\
Action Refinement & $0.330$~ms & $0.040$~ms \\
Global Explanation & $0.610$~ms & $0.080$~ms \\
\bottomrule
\end{tabular}%
}%
\vspace*{-1.5ex}%
\end{table}

\vspace*{-1ex}
\subsection{Improving Agent Performance via Action Refinement}
\label{subsec:action-refinement-results}
\vspace*{-1ex}

\begin{figure}[t]
\centering
\includegraphics[width=0.9\columnwidth,keepaspectratio]{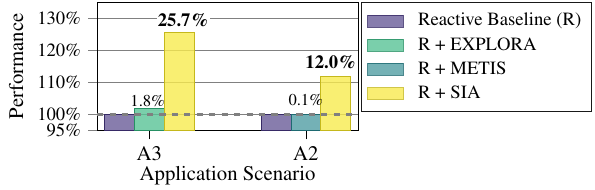}
\vspace{-2ex}
\caption{Performance gains via \toolnamex's action refinement.}
\label{fig:action_refinement_result}
\vspace{-1.5ex}
\end{figure}

To address \textit{\ref{rq3}}, we evaluate \toolname's Action Refinement module, which gives reactive agents forecast-awareness without retraining. This module yields considerable gains, as shown in Figure~\ref{fig:action_refinement_result}. Applying it to the \ac{ran} slicing agent (\AthreeRSIA) achieves a $25.7$\% cumulative reward boost, substantially outperforming EXPLORA's $1.8$\% gain. Similarly, the refined \ac{mimo} agent (\AtwoRSIA) provides a $12.0$\% reward increase, far exceeding the $0.1$\% from Metis.

These gains are possible because \toolname's bounded symbolic state space and efficient \ac{kg} queries enable rapidly identifying the historically optimal action for a forecasted state transition. The refiner overrides the agent's decision if this action's expected reward exceeds the original by a configurable threshold, $\tau$ (set to 3\% in our experiments), as detailed in Algorithm~\ref{alg:action-refiner}. This override check completes in just $0.33 \pm 0.04$\,ms (see \S\ref{subsec:performance-eval}), meeting the O-RAN near-RT RIC's timing budget and offering a practical path to proactive control in production networks where retraining is often prohibitive.

These gains are robust to moderate forecast inaccuracies. Because the refiner operates on symbolic categories, performance is robust to forecast errors unless they are large enough to shift a \ac{kpi}'s value across a percentile boundary, a resilience mechanism inherent to \toolname's design.

% = = = = = = = = = = = = = = = = = =
\vspace*{-0.75ex}
\section{Discussion and Limitations}
\label{sec:disc}
\vspace*{-0.75ex}
% = = = = = = = = = = = = = = = = = =

\noindent\textbf{Discussion.}\;
\toolname offers a practical methodology for interpreting anticipatory \ac{drl}. Its utility is shown by its generalizability across diverse use cases and its ability to generate real-time explanations using our novel \ac{is} metric. Its use of per-\ac{kpi} \acp{kg} avoids the state-explosion problem of monolithic approaches, ensuring its overhead and memory footprint remain scalable for production. We show \toolname's insights are actionable: they expose actual design flaws, such as the \ac{mimo} agent's reward bias, and guide an \ac{abr} agent redesign that increases bitrate by 9\%. Moreover, its Action Refinement module boosts \ac{ran}-slicing agent's reward by 25\% without retraining.

\noindent\textbf{Limitations.}\;
Through an ablation analysis of \toolnamex’s operating boundaries, we identified three key limitations and deployment considerations: 
(i) both \toolnamex's explanations and the Action Refiner's suggestions (see~\S\ref{subsec:action-refinement-results}) are affected by forecast accuracy. However, \toolnamex’s symbolic foundation provides resilience, as performance is largely unaffected unless forecast errors are large enough to push a \ac{kpi} across a category boundary. While this implies higher sensitivity for stable metrics with narrow percentile bands, such \acp{kpi} are typically forecasted with high accuracy, which mitigates the risk. 
(ii) \toolname exhibits a cold-start phase because the per-\ac{kpi} \acp{kg} are initially sparse. In practice, this can be mitigated by pre-populating the \acp{kg} from offline traces. 
(iii) The framework is sensitive to the Symbolizer's configuration. We found the number of categories is a critical hyperparameter: using fewer than three produces overly generic insights, while more than seven prolongs the cold-start period. An odd number of categories (e.g., five) is preferable, as it provides a distinct middle category (i.e., \lstinline!Medium!). These findings led to our default of five and three categories for a \ac{kpi}'s value and trend, respectively (see~\S\ref{subsec:symbolizer}). Together, these aspects define the practical conditions for \toolnamex's successful deployment.

% = = = = = = = = = = = = = = = = = =
\vspace*{-0.75ex}
\section{Conclusions}
\label{sec:concl}
\vspace*{-0.75ex}
% = = = = = = = = = = = = = = = = = =
This paper presented a new paradigm for interpreting anticipatory \ac{drl} agents by introducing \toolname, a symbolic framework that brings transparency and trust to their operation in mobile networking. By leveraging scalable per-\ac{kpi} knowledge graphs and a novel \ac{is} metric, \toolname delivers real-time, actionable insights that are beyond the reach of existing methods. Our evaluations demonstrated that these insights are not merely diagnostic; they enable both targeted agent redesigns and automated performance enhancements, boosting key network metrics by up to 25\%. Ultimately, by making proactive control understandable and tunable, \toolname lowers a critical barrier to its adoption in next-generation networks.

\vspace{-1ex}
% = = = = = = = = = = = = = = = = = =
\section*{Acknowledgments}
% = = = = = = = = = = = = = = = = = =
\vspace{-0.5em}
This work is partially supported by bRAIN project PID2021-128250NB-I00 funded by MCIN/AEI/10.13039/501100011033/ and the European Union ERDF ``A way of making Europe''; by Agile-6G Project PID2024-163089NB-I00 funded by MICIU/AEI/10.13039/501100011033; C. Fiandrino is a Ram\'on y Cajal awardee (RYC2022-036375-I), funded by MCIU/AEI/10.13039/501100011033 and the ESF+. This work is also supported by U.S. NSF under grants CNS-2112471 and CNS-2434081, and by OUSD(R\&E) through ARL CA W911NF-24-2-0065. The views and conclusions contained in this document are those of the authors and should not be interpreted as representing the official policies, either expressed or implied, of the Army Research Laboratory or the U.S. Government. The U.S. Government is authorized to reproduce and distribute reprints for Government purposes notwithstanding any copyright notation herein.

\balance
\bibliographystyle{IEEEtran}
\bibliography{sample-base}

\end{document}